\documentclass[twocolumn]{aastex62}

\graphicspath{{./}{figures/}}

\accepted{December, 2018}
\submitjournal{ApJL}

\shorttitle{Expanding Young Star Populations in Perseus}
\shortauthors{Rom\'an-Z\'u\~niga et al.}

\begin{document}

\title{Evidence of Hubble flow-like motion of young stellar populations away from the Perseus arm}

\correspondingauthor{Carlos G. Rom\'an-Z\'u\~niga}
\email{croman@astro.unam.mx}

\author[0000-0001-8600-4798]{Carlos G. Rom\'an-Z\'u\~niga}
\affil{Instituto de Astronom\'ia, Universidad Nacional Aut\'onoma de M\'exico, Unidad Acad\'emica en Ensenada,  Ensenada 22860 Mexico}
\affil{Programa de Estancias de Investigaci\'on (PREI), DGAPA-UNAM, Mexico}

\author[0000-0002-1379-4204]{Alexandre Roman-Lopes}
\affiliation{Departamento de F\'isica y Astronom\'ia, Universidad La Serena, La Serena, Chile}
\affil{Programa de Estancias de Investigaci\'on (PREI), DGAPA-UNAM, Mexico}

\author[0000-0002-0506-9854]{Mauricio Tapia }
\affil{Instituto de Astronom\'ia, Universidad Nacional Aut\'onoma de M\'exico, Unidad Acad\'emica en Ensenada,  Ensenada 22860 Mexico}
\affil{Programa de Estancias de Investigaci\'on (PREI), DGAPA-UNAM, Mexico}

\author[0000-0001-9797-5661]{Jes\'us Hern\'andez}
\affil{Instituto de Astronom\'ia, Universidad Nacional Aut\'onoma de M\'exico, Unidad Acad\'emica en Ensenada,  Ensenada 22860 Mexico}

\author[0000-0000-0000-0000]{Valeria Ram\'irez-Preciado}
\affil{Instituto de Astronom\'ia, Universidad Nacional Aut\'onoma de M\'exico, Unidad Acad\'emica en Ensenada,  Ensenada 22860 Mexico}



\begin{abstract}

In this letter we present evidence of coherent outward motion of a sample of young stars ($t<$30 Myr) in the Perseus Arm, whose apparent origin is located in the vicinity of the W3/W4/W5 complex. Using astrometric and photometric data from the Gaia DR2 catalog of an 8$^\circ$ radius field centered near W4, we selected a sample of young, intermediate to high-mass star candidates. The sample is limited to sources with parallax uncertainties below 20\% and Bayesian distance estimates within 1800 and 3100 pc. The selection includes embedded stellar populations as well as young open clusters. Projected velocities derived from perspective-corrected proper motions clearly suggest that the young star population emerge from the Perseus arm, with a possible convergence zone near W3/W4/W5 region, tracing a front that expands away at a rate of about $15~{\rm km~s}^{-1}~{\rm kpc}^{-1}$.

\end{abstract}



\section{Introduction} \label{sec:intro}

Star clusters are formed within Giant Molecular Clouds (GMC) with a large diversity, likely related to the early organization of the clouds and environmental factors
\citep{lada:2003,longmore:2014}.  In a recent review, \citet{goullermis18} (and references therein) discuss the case of gravitationally unbound young stellar systems, representing GMC complex-scaled multiplex associations that may be considered to be the open end of an intricate, hierarchical process that scales down to the formation of compact, bound clusters. In that scenario, large GMC complexes may produce a large number of star clusters born with a variety of energy distributions \citep{blaauw64} that, in most cases: a) expand and disperse, b) can be traced through their massive (OB) members, and c) may contain crucial information about their formation process. In general, the expansion and dispersal of young star systems may be related to complex dynamical interactions between clusters and sub-clusters during their formation process, and also to the rapid removal of gas due to massive star winds and more violent events like Wolf-Rayet outbursts or supernova explosions. The current availability of precise astrometric parameters (distances, proper motions) from the Gaia satellite Data Release 2 \citep[hereafter GDR2;\ see][]{gaia:2016,gaia:2018} allows the exploration of star kinematics in this context, and it has provided evidence of coherent motion and expansion of stellar populations in certain stellar systems, from star forming complexes \citep[e.g.][]{kounkel:2018,kuhn18} to more evolved young cluster associations \citep[e.g.][]{wright18} and OB associations with driven expansion \citep[e.g.][]{cantat18}.

\begin{figure}
\includegraphics[width=3.1in]{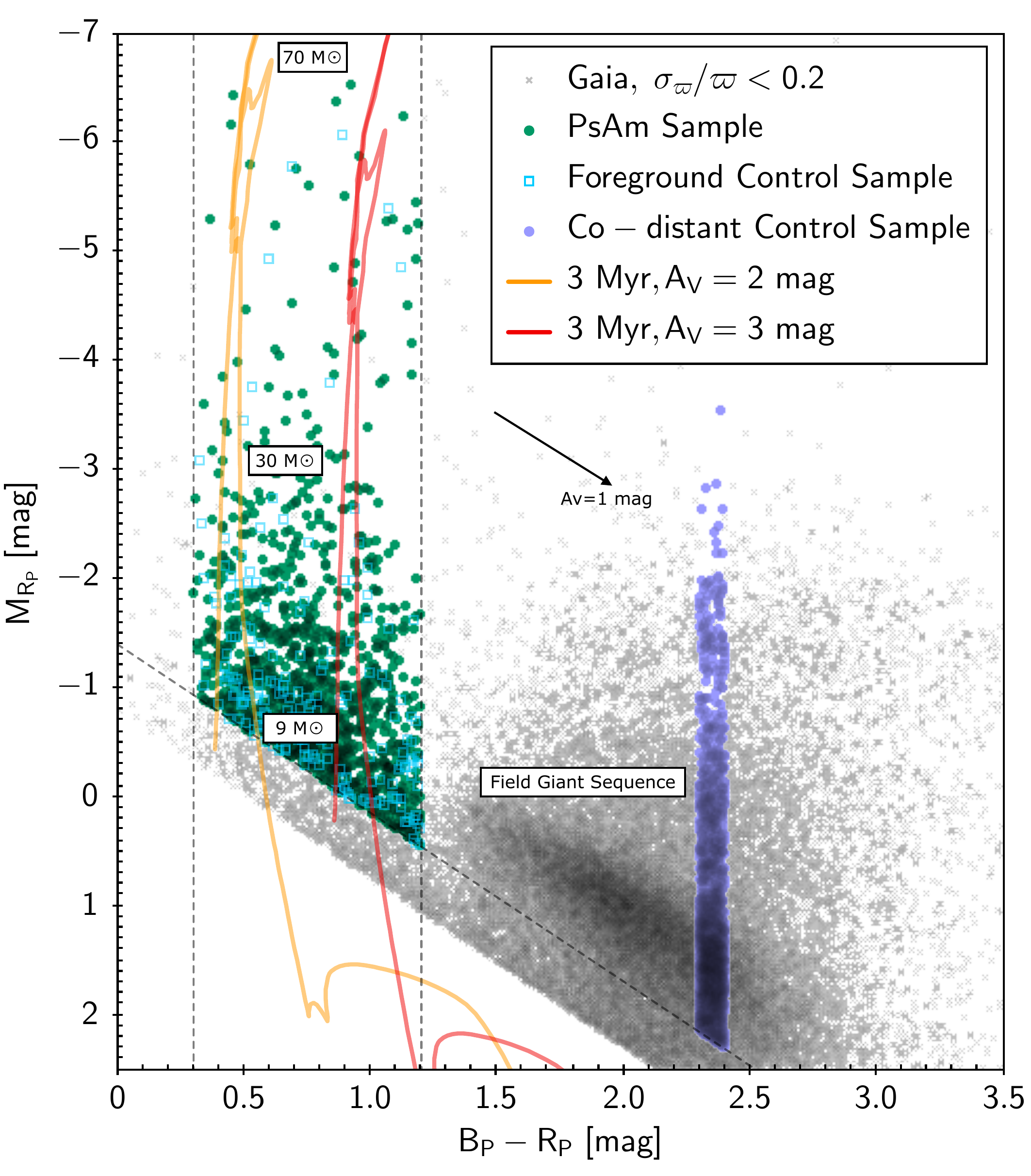}
\caption{Gaia color-absolute Magnitude diagram showing the selection of a sample containing OB stars. Gray symbols are GAIA sources in the main catalog with parallax errors under 20 percent. Green symbols indicate the PsAm sample. Yellow and cyan symbols show the control samples. The color/brightness constraints ($\S$\ref{sec:sample}) are indicated with dotted lines. The solid color lines are 3 Myr isochrones from the PARSEC 1.2s models \citep{bressan:2012,tang:2014}.  \label{fig:cmdiag}}
\end{figure}

The goal of this letter is to report a discovery about the global picture of star cluster evolution in a region of the Perseus arm (hereafter PsAm), around the W3/W4/W5 complex (hereafter W345). W345, located on the Galactic plane, is a well studied massive star forming region, containing two giant HII regions (W4 and W5), a massive molecular ridge with active formation (W3) and several embedded star clusters \citep[e.g.][]{carpenter:2000,koenig:2008,roman:2015,jose:2016,sung:2017}, possibly hosting a large number of OB stars. Around W345, this region of the PsAm hosts several OB associations like Per OB1 that contains the open clusters NGC 884 and 869, and the Cassiopeia OB8 that includes the cluster NGC 663.  Here we show that a sample of intermediate to high mass young stars in W345 and nearby associations, traces what looks like a coherent, expanding front of young ($t<30$ Myr) stars and clusters, likely born at the PsAm close to the W345 region.

\section{Data Sample Selection and Analysis} \label{sec:sample}

Our main goal is to study the kinematics of intermediate to massive (OB) stars as tracers of the young stellar population in W3/W4/W5 and its surroundings. Using the Gaia Table Access Protocol (TAP) services and catalog extraction tool at Astronomisches Rechen-Institut\footnote{http://gaia.ari.uni-heidelberg.de/}, we retrieved a catalog comprising all DR2 entries within an 8 degree-radius cone centered on the W345 complex at $(\alpha,\delta)=(40\arcdeg,61\arcdeg)$\footnote{We confirmed that a larger opening does not alter our results significantly.}. The catalog contains a full set of photometric ($G$, $B_p$ and $R_p$ magnitudes) and astrometric (proper motions, $\mu_P$, parallaxes, $\varpi$ and parallax errors, $\sigma_\varpi$) parameters. This cone extends roughly from 250 to 420 pc (depending on distance) in each direction from the W345 complex, covering the projected width of the PsAm and providing a good sampling of the populations formed at GMC complexes in the region. 

The TAP tool also provides distance estimates for each star by using a Galactic stellar density model prior that varies with each position $(l,b)$ \citep{BJ:2018} and provides  confidence intervals, defined by lower and upper limits, $d_{low}$ and $d_{high}$. We extracted sources within $1.8<(d/kpc)<3.1$, which we define as a ``full" sample expected to contain sources across the entire girth of the PsAm. We limited the sample to sources with $\sigma_\varpi/\varpi<0.2$ parallax accuracy\footnote{a tighter constraint of $\sigma_\varpi/\varpi\leq 0.1$ does not alter significantly our results} and contained within a distance range defined by $(d_{high}-d_{low})<1.2d+20$. As shown in Figure \ref{fig:cmdiag}, using the $M_{R_P}\mathrm{vs.\ }{B_P-R_P}$ Gaia color-absolute magnitude diagram, we selected a locus containing sources brighter than $1.547\times(B_P-R_P)-1.2$ \citep[the slope follows the reddening vector using the extinction coefficients from][]{evans:2018} and colors within $0.3<B_P-R_P<1.2$ mag. This locus minimizes contamination from field giant and dwarf stars, and is expected to contain mostly bright intermediate to massive members of young stellar groups with average extinctions within $A_V=$1 and 3 mag; \textbf{still, we estimate around 10\% of contamination in this sample due to field stars with anomalous colors, with a distribution affected by patchy extinction}. Finally, we also restricted the projected raw velocity moduli to $\sqrt{v_l^2 + v_b^2}=|v|<60 \mathrm{~km~s^{-1}}$ . The total number of sources in this restricted PsAm sample is 1100 sources.

\begin{figure*}
\begin{center}
\includegraphics[width=6.5in]{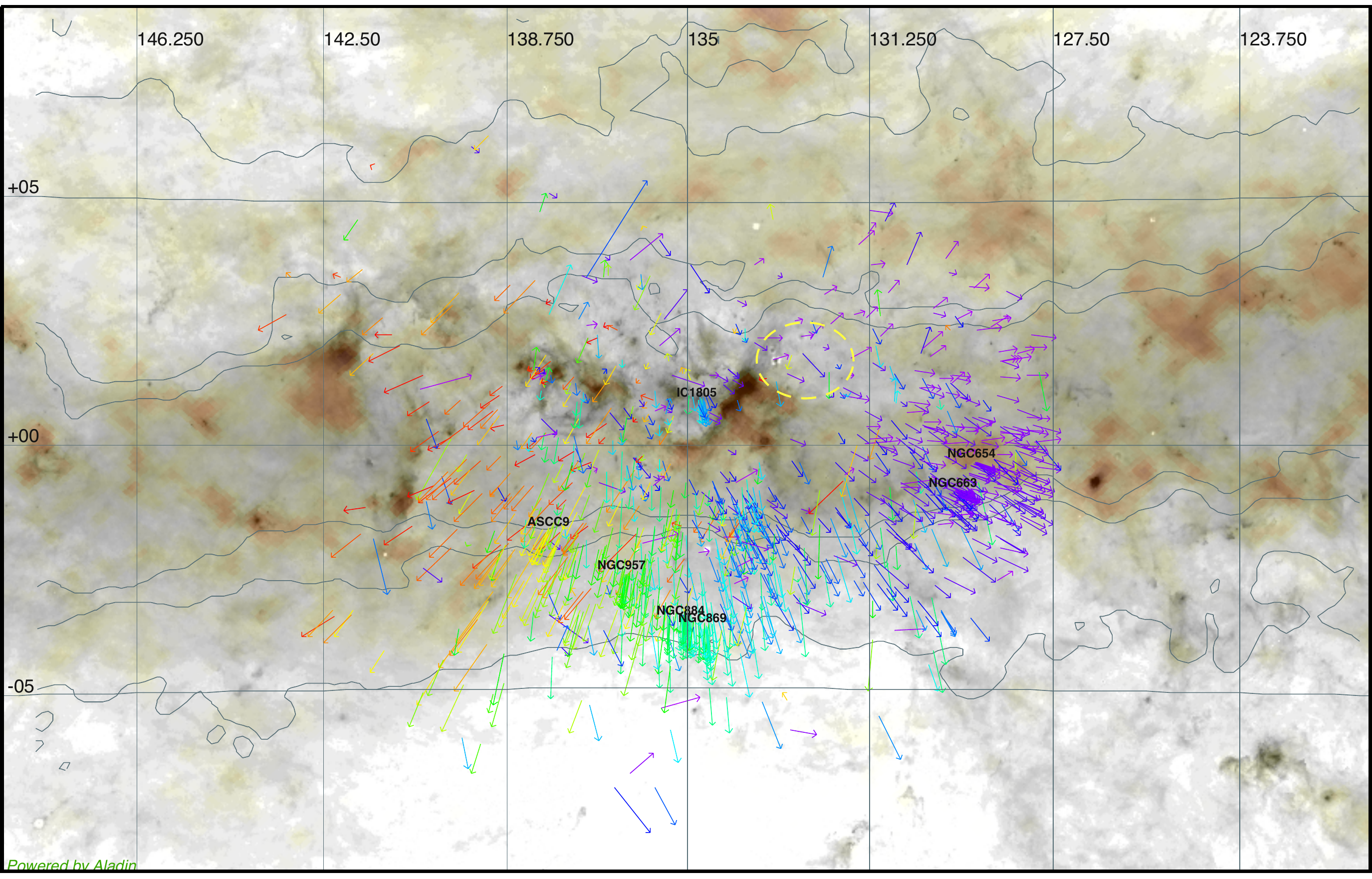}
\caption{ \textit{Top:} Perspective corrected, projected velocity map for the full sample, in Galactic coordinates. The background in gray scale is a WISE WSSA 12$\mu m$ infrared emission map. The orange-red hue indicates regions of strong CO emission from the composite survey of \citet{dame:2001}. Notice that most of the CO emission above $b\sim 3^\circ$ corresponds to clouds in the Local Arm, not Perseus. The contours indicate HI column density \citep[EBHIS+GASS surveys][]{HI:2016} in levels of 50 to 200 cm$^{-2}$/healpix. The arrows indicate projected velocities scaled to units of $500\mathrm{~pc~Myr^{-1}}$ for visualization purposes. The colors of the arrows in this map are parametrized by $\cos{\theta}$. The region delimited by the yellow dashed-line indicates the approximate extension of the HB-3 supernova remnant near W3. Young open clusters in the region are indicated with labels. \label{fig:map}}
\end{center}
\end{figure*}

From the distances and proper motions, we calculated projected velocities over the plane of the sky on galactic coordinates as $v_{l,b}=4.74d\mu_{l,b}$, where $\mu_l$ and $\mu_b$ are directly transformed from the $\mu_\alpha$ and $\mu_\delta$ values provided in the GDR2\footnote{all proper motions and velocities are expressed units of mas/yr and km/s, respectively}. The angular size of the region of study is quite large, thus requiring a correction by perspective expansion/contraction \citep{brown97} to take into account the effect of the motion perpendicular to the plane of the sky. We followed the prescription of \citet{helmi18}, by first defining an expansion center  at $(l,b)=(137.8\arcdeg,1.3\arcdeg)$ (see also $\S$\ref{sec:results}), located near the young cluster IC 1805 in W4. We then derived a new set of orthographically projected positions ($x_p, y_p$), and proper motions ($\mu_{x_p}, \mu_{y_p}$),  relative to that center. Assuming that rotation and inclination effects in the PsAm are negligible, the variation of the projected proper motions as a function of the projected positions, to first order, will only keep the bulk motion component perpendicular to the line of sight: ${\partial \mu_x} /{\partial x}\approx {\partial \mu_y} /{\partial y} \approx -v_z$. This way,  a linear fit to either of the $\mu_{x_p}$ vs $x_p$ or $\mu_{y_p}$ vs $y_p$ plots provides, to first order, a direct estimate of the required perspective correction. In our case, the resultant correction is $c_{\mu_{l,b}}\approx 5.4(x_p,y_p)\mathrm{~mas~yr^{-1}}$, applied to the orthographically projected positions, which we finally subtracted from the GDR2 proper motions. This accounts for the required perspective correction by the unknown line-of-sight motion of each star. 

\textbf{Now, it is important to consider that without radial velocity data, it is impossible to determine whether the observed expansion is due to purely perspective or purely physical effects. In order to quantify this, we followed \citet{blaauw64} and \citet{wright18} (their equations 3 and 4): for a system in linear expansion with an age $\tau$, it is possible to predict the required radial velocities as the sum of the star velocities measured from the system barycenter and an expansion age term\footnote{plus a redshift/blueshift correction, which is negligible in our case},  $\kappa d=(1.0227\tau)^{-1}d$. Using a Monte Carlo routine, we found that for stars with $1<\tau<50$ Myr (see section \ref{sec:results}), the expected distribution for $\kappa d$ has a well defined peak around 100 km/s. We are able to reproduce well that distribution by providing our sample with radial velocities values in the range $-15<v_R<-65$ km/s, i.e. around the LSR velocity for the W3/W4/W5 clouds \citep[e.g.][]{heyer98,bieging11}. We failed to emulate the $\kappa d$ distribution when we increased the radial velocity interval to over 100 km/s. Radial velocity corrections so large are physically unreasonable as a solely perspective effect. Therefore, we can safely consider our sample to present a physical expansion}.

\begin{figure}
\includegraphics[width=3.2in]{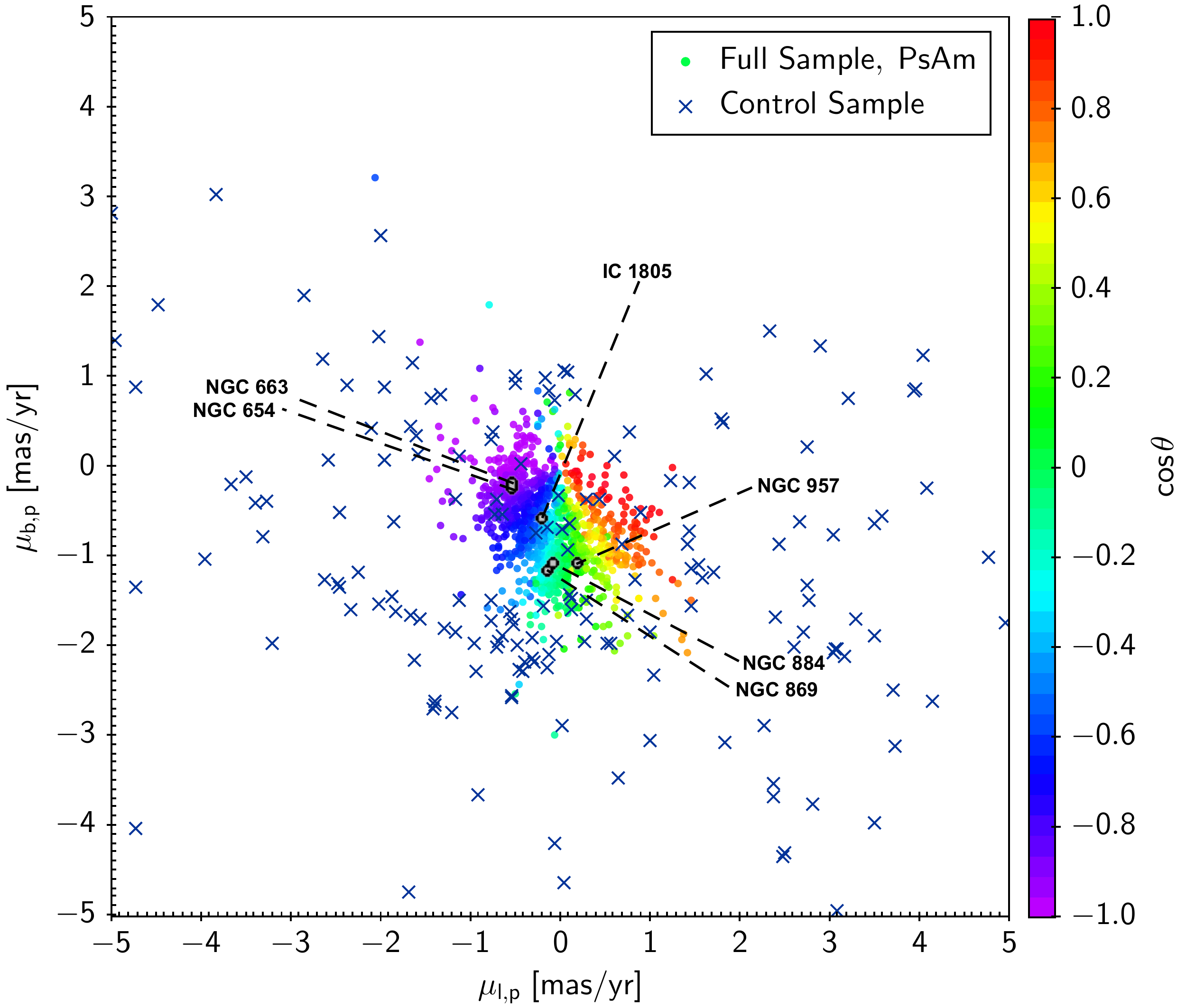}
\caption{Perspective corrected, proper motions for the full sample (color parametrized by $\cos{\theta}$) and the control sample (blue X symbols). The average values for 
cluster member candidates are indicated with the black dots and labels. \label{fig:dotvector}}
\end{figure}

\begin{figure}
\includegraphics[width=3.2in]{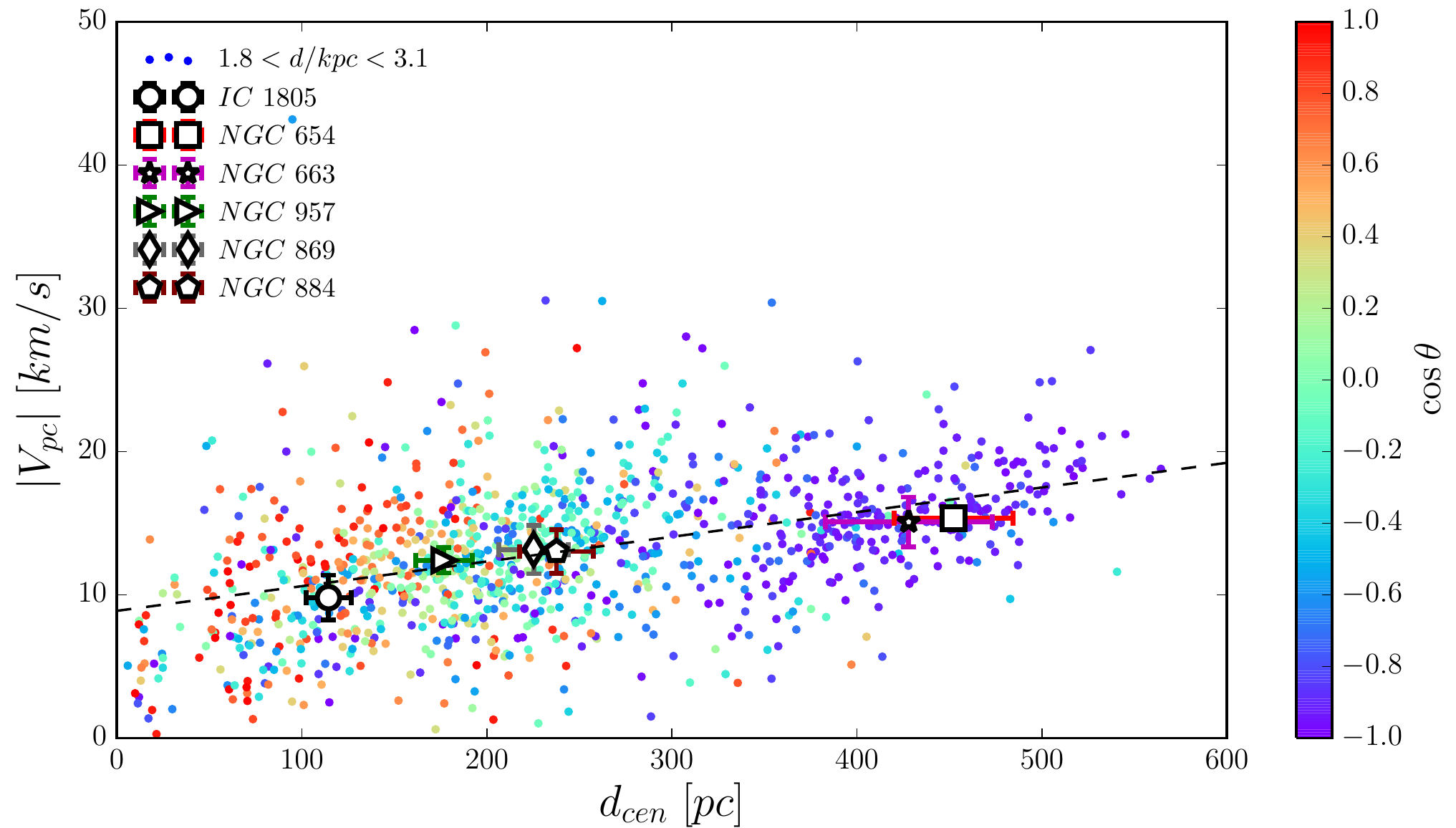}
\includegraphics[width=3.2in]{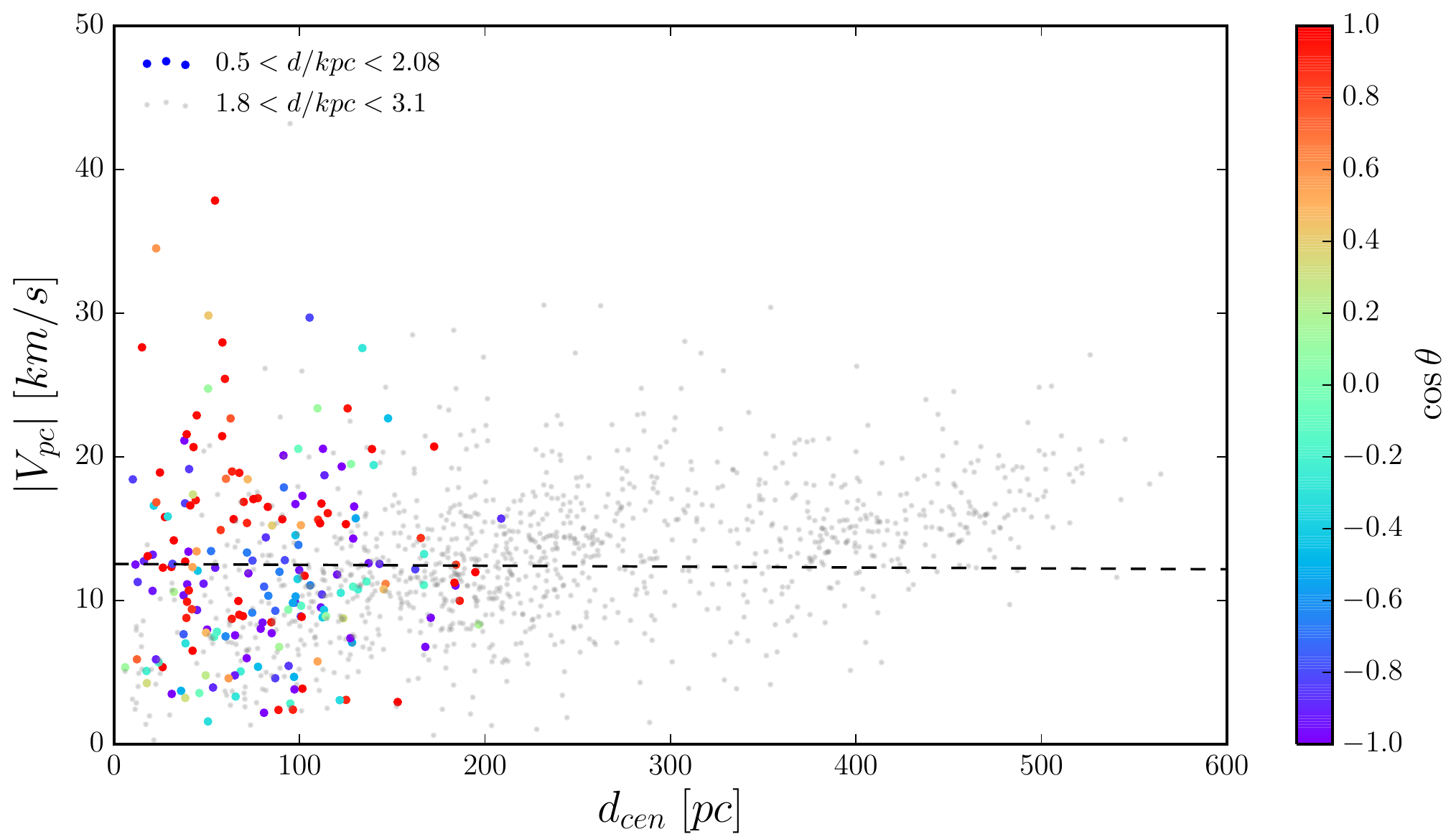}
\caption{Perspective corrected, projected velocity moduli $|v|$ as a function of projected distance, $d_{cen}$ from an estimated convergence point (l=137.8 deg, b=1.3 deg). The top and bottom panels show the result for the PsAm and the control sample, respectively. The positions of known young open clusters are shown as large, polygon shaped symbols with error bars. \label{fig:vdcen}}
\end{figure}

\textbf{Finally, we applied the same constraints to two control samples within the same cone: one (foreground control sample) limited to sources with distances $d<1300$\ pc (i.e. unrelated to the Perseus arm) and a second one (co-distant control sample), containing stars in the same distance range but sampling an older population dominated by field giants within $2.3<(B_P-R_P)<2.4$. The resultant samples contain 193 and 2468 stars respectively. The purpose of these control samples is to apply a similar analysis to them, and assure that the observed behavior (see section \ref{sec:results}) of the young massive candidates in the Perseus arm is not some data-related effect (e.g. present in the whole DR2 Gaia sample in this direction of the Galaxy).}

\section{Results} \label{sec:results}

In Figure \ref{fig:map} we show the perspective-corrected, projected velocities $v_{l,pc},v_{b,pc}$ (scaled to units of $500$ pc/Myr) for the DR2 Gaia stars present in the full distance sample, over a map that traces star forming regions, as well as the distribution of molecular and neutral gas near W345, using 12 $\mu$m emission, as well as HI and CO integrated intensity. As a visualization aid, we use an orientation parameter by calculating the dot product between the corrected projected velocity, $v_{l,pc}$, and a purely horizontal unit proper motion vector $v_{l,pc}=(1,0)$, as $\cos{\theta}=v_{l,pc}/|v_{l,pc}|$. \textit{The most striking feature in this map is the high degree of coherence in the orientation of the vectors on the plane of the sky, pointing away from W345, increasing in magnitude with distance}. On the other hand, the selected sample covers the current star forming regions in W345, and highlights several source overdensities that correspond to well known young open clusters (see also Figure \ref{fig:vdcen}). For instance, the overdensity near $(l,b)=(135,-0.4)$ corresponds to NGC 884 and NGC 869 in the Per OB1 association, also known as $h+\chi$ Persei \citep[10-13 Myr old, $d=2.5\pm 0.2$ kpc][]{currie:2010}. We can also distinguish two clear groups in the westernmost edge corresponding to NGC 663 and NGC 654 of the Cas OB8 association \citep[20 Myr old, $d=2.4\pm 0.2$ kpc][]{tapia:1991,pandey:2005}\footnote{Photometric distances. The GDR2 parallaxes suggest larger distances, near 3 kpc, for these two clusters}. Another very important feature that emerges from Fig. \ref{fig:map} is that for the perspective corrected motions for the control sample, we cannot identify any grouping, and the vectors present a net null trend, with the majority of the sources moving in directions parallel to the Galactic disk, thus confirming that the above effect is actually associated to the Perseus arm population. 

In Figure \ref{fig:dotvector} we show the perspective-corrected proper motions (a dot-vector plot) for the full and the control sample. We clearly see that while the PsAm stars form a tight, well defined structure with several overdensities (that correspond well with the larger groups), the control sample sources appear significantly dispersed, indicating very little or no coherence. In the first panel of the Figure set \ref{fig:vdcen} we show the perspective-corrected, projected velocity modulus $|v|$ as a function of the projected distance, $d_{cen}$, from a convergence area centered at $(l,b)=(137.8,1.3)$, which was estimated by projecting backwards the $v_l,v_b$ vectors of candidate members of known open clusters, (defined as sources within a small vicinity around published centers and distances and limited to $\sigma _\varpi/\varpi<0.1$). \textit{We can see how the corrected velocity moduli for most of the sources are in the range of 10-20 km/s, which is significantly larger than the typical expansion velocities for unbound groups in other systems} \citep[e.g.][]{cantat18,kuhn18,wright18}. It is clear that the young stellar population is actually expanding away from the Galactic plane: a linear fit to the data suggests a (projected) Hubble-type expansion of about 15.4 $\mathrm{km\ s^{-1}\ kpc^{-1}}$. The additional panel of the Figure set \ref{fig:vdcen} shows the equivalent plot for the control sample, where we confirm a null expansion trend, with a correlation coefficient near zero. \textbf{The results for the co-distant control sample are equivalent, with a null expansion trend and little or no coherence in the dot-vector plot}.

\section{Discussion and Summary} \label{sec:discussion}

The analysis of the Gaia DR2 data described above, reveal a scenario where young stellar populations are moving away from a location in the Perseus arm. Tracing back the $(v_l,v_b)$ vectors points to a convergent region near the W345 complex that currently hosts embedded populations as young as 1.5 Myr (likely the most recent star formation episode in the region). A satisfactory explanation of the effect and its consequences for currently accepted Galactic star formation scenarios is beyond the scope of this letter, as more precise data would be required. For instance, the Gaia DR2 parallaxes provide distances that are still uncertain in the $10^2$ pc scale for the Perseus arm, impeding an accurate analysis of the distribution of young stellar populations within this part of the Perseus arm. Also, the lack of radial velocity measurements for most of the stars in our sample, prevents a full 6-dimensional position-velocity analysis at this Galactic region. From the observed expanding motion and the available theories in the literature, we propose three possible scenarios:

1. The presence of the HB-3 SN remnant near W3, along with the presence of tens of OB stars in the W345 complex (see top panel Fig.  \ref{fig:map}), suggest that supernova explosions could be one possible driving mechanism in the region, which could transfer the mechanical energy to the cluster-gas systems. Indeed, it is known that a single SN can imprint momentum of the order of $10^5\mathrm{\ M_\odot \ km/s}$ \citep{kim:2015}, providing up to $10^{50}$ ergs in 10 Myr \citep{thornton:1998}. Super-bubble type expansions via SN often pierce and surpass the disk, depending on the homogeneity of the local medium \citep{mordecai:1989}. Also, most of the mechanical energy from SN events and massive star winds is carried by ionizing radiation, which could provide impulse for the stellar system expansion. Gaia DR2 data could be providing evidence of a champagne flow type expansion of gas layers \citep{tenorio:1979}, with the stars moving \textit{along with the gas.}
   
2. Recently, \citet{peek:2018} presented detailed gas velocity maps and compared them to the Stationary Density Wave model of the Milky Way spiral structure \citep{linshu:1964}, particularly near the Perseus arm. The study suggests that a spiral arm can be fed from one side by diffuse gas causing a net outward flow toward its trailing edge. The main effect is a concentration of the dense gas where flows converge, and a dissipation of those centers after star formation, forming a divergence of the velocity field. This mechanism can provide a direct explanation of the observed effect in our sample. Moreover, if gas and stars are expelled away from the arm, aided by this mechanism, then we should expect a net torque in the arm. If this is true, our study presents actual evidence of the action of spiral density waves near the co-rotation radius in the Perseus arm.

3. Another possibility is that all the groups we are considering are all part of a much larger association of clusters that is expanding away as a massive unbound young stellar system with an age spread of about 30 Myr (considering the estimated ages of the older clusters in the sample). The driving mechanisms behind the expansion could be directly related to the star formation process, along with gas dispersal and the dynamic evolution of the system, similar to what is observed in star forming complexes like Orion, Vela OB2 or Scorpio-Centaurus. However, it would be necessary to determine how to adjust such proposed scenarios to the spiral arm width scale we report in this paper.
  
While eloquent, our analysis is just a starting point. Future work requires to search for this effect in other massive star forming regions in the nearby spiral arms. Large scale spectroscopy projects like APOGEE-2 will provide precise radial velocity information at least near W345 \citep[see\ ][]{gail:2017}, allowing for a more complete vision, specially as the precision of Gaia data increases in the following releases.

\acknowledgements

We thank an anonymous referee for providing crucial comments and suggestions that greatly improved the manuscript. We thank Luis Aguilar for fruitful discussions and advice. VRP acknowledges support from a CONACYT/UNAM scholarship. CRZ, VRP, MT and JH acknowledge support from UNAM-DGAPA-PAPIIT grants IN108117, IN104316 and IA103017, Mexico. ARL acknowledges support from program FONDECYT 1170476, Chile and a PREI-DGAPA-UNAM academic exchange scholarship.

This work has made use of data from the European Space Agency (ESA)
mission {\it Gaia} (\url{https://www.cosmos.esa.int/gaia}), processed by
the {\it Gaia} Data Processing and Analysis Consortium (DPAC,
\url{https://www.cosmos.esa.int/web/gaia/dpac/consortium}). Funding
for the DPAC is provided by the institutions participating in the {\it Gaia} Multilateral Agreement. 
\vspace{5mm}
\facilities{Gaia}
\software{Aladin \citep{aladin}, TOPCAT \citep{topcat}, AstroPy \citep{astropy1,astropy2}}



\end{document}